\documentclass[useAMS,usenatbib,referee]{biom}
\usepackage{graphicx}

\let\theoremstyle\relax

\usepackage{amsmath,amsfonts,amsthm,bm} 
\usepackage{hyperref}
\usepackage{enumitem}
\usepackage{standalone}
\usepackage{color}
\usepackage[flushleft]{threeparttable}

\usepackage{algorithm,algpseudocode}
\usepackage{subcaption}
\usepackage[makeroom]{cancel}

\def\bSig\mathbf{\Sigma}

\title[Network-based Topic Interaction Map for Big Data Mining]{Network-based Topic Interaction Map for Big Data Mining of COVID-19 Biomedical Literature}

\author{Yeseul Jeon$^{1,2}$, 
Dongjun Chung$^{3}$, 
Jina Park$^{1,2}$,
and Ick Hoon Jin$^{1,2,*}$\email{ijin@yonsei.ac.kr} \\
$^{1}$Department of Statistics and Data Science, Yonsei University, Seoul, South Korea\\
$^{2}$Department of Applied Statistics, Yonsei University, Seoul, South Korea\\
$^{3}$Department of Biomedical Informatics, The Ohio State University, Columbus, Ohio, U.S.A}

\begin{document}

\label{firstpage}

\begin{abstract}
Since the emergence of the worldwide pandemic of COVID-19, relevant research has been published at a dazzling pace, which yields an abundant amount of big data in biomedical literature. Due to the high volum of relevant literature, it is practically impossible to follow up the research manually. Topic modeling is a well-known unsupervised learning that aims to reveal latent topics from text data. In this paper, we propose a novel analytical framework for estimating topic interactions and effective visualization to improve topics' relationships. We first estimate topic-word distributions using the biterm topic model and estimate the topics' interaction based on the word distribution using the latent space item response model. We mapped these latent topics onto networks to visualize relationships among the topics. Moreover, in the proposed approach, we developed a score that is helpful in selecting meaningful words that characterize the topic. We figure out how topics are related by looking at how their relationships change. We do this with a "trajectory plot" that is made with different levels of word richness. These findings provide a thoroughly mined and intuitive representation of relationships between topics related to a specific research area. The application of this proposed framework to the PubMed literature demonstrates utility of our approach in understanding of the topic composition related to COVID-19 studies in the stage of its emergence.
\end{abstract}


\maketitle

\section{Introduction}
\label{s:intro}

Since December 2019, there has been a worldwide pandemic known as COVID-19, which has resulted in over 636 million illnesses and over 6 million fatalities worldwide. Tremendous efforts to understand and overcome this fatal virus have led to a rich amount of biomedical literature about COVID-19. 
As of November 2022, searching with the term ``COVID-19'' in the PubMed biomedical literature database (\url{https://pubmed.ncbi.nlm.nih.gov}) resulted in more than 316,000 publications. Following the 1918 flu pandemic, COVID-19 is the second rapidly spreading pandemic that has yielded big data in epidemiology. Many researchers have explored COVID-19, which covers a wide range of topics, including general description, mechanism, transmission, diagnosis, treatment, prevention, case report, and forecasting, among others, as profiled and categorized in LitCovid, a biomedical literature database dedicated to COVID-19 \citep{RN12503}. In order to facilitate an understanding of relevant mechanisms, it is critical to follow up and digest these publications that are being published at a fast phase. However, given the rich volume of COVID-19 literature and its rapid publication phase, it is practically impossible for biomedical experts to trace all of this literature manually in real-time. 


In the text mining area, topic modeling is the most well-known approach for this purpose. It essentially extracts topics based on the words' distribution in each document and various algorithms have been proposed for topic modeling. \citet{blei2003latent} made an important step in this direction and proposed a generative probabilistic model for word sets, called the Latent Dirichlet Allocation (LDA). LDA assumes each document is structured in the context of topics while words are allocated to these topics. The topics are regarded as latent variables and characterized by words with probabilities assigned to each topic. From the perspective of biomedical literature text mining, \citet{liu2016overview} applied topic modeling to biomedical literature and showed that topic modeling could be a practical solution for bioinformatics research.

Since COVID-19 was a worldwide pandemic that spread at an unprecedented rate, some articles that examine COVID-19 from a variety of perspectives have been published with only an abstract. Abstracts of the COVID-19 literature are categorized as short text. Unfortunately, with short texts, conventional topic models including LDA often suffer from poor performance due to the sparsity problem. {\citet{yan2013biterm} made important progress in the modeling of short text data, the so-called Biterm Topic Model (BTM). Unlike the LDA, BTM replaced words with bi-terms, where a bi-term is defined as a set of two words occurring in the same document. This approach attempted to compensate for the lack of words in a short text by pairing two words, thereby creating more words in the document. This is based on observation that if two words are mentioned together, they are more likely to belong to the same topic. Some studies that applied BTM to short text data include micro-blog \citep{li2016micro}, large-scale short texts, and tweets for clustering and classification \citep{chen2017word}.


It is important to note that it is hard to find the data that contain independent topics towards one subject in the real world. Rather, instead, multiple topics are often inter-correlated. Therefore, to tackle this practical issue, there have been attempts to model interactions among topics by modeling the correlated structure within the topic model or combining the topic model with statistical models. For instance, \citet{blei2007correlated} regarded the topic of correlation as a structure of heterogeneity. To capture the heterogeneity of topics, they suggest the Correlated Topic Model (CTM), which models topic proportions to exhibit correlation through the logistic normal distribution within LDA. They validated their performance by applying CTM to the articles from the journal Science. On the other hand, \citet{rusch2013model} combined LDA with decision trees to interpret the topic relationships from Afghanistan war logs. It classified the topics using tree structures, in which helped to understand the different circumstances in the Afghanistan war. However, these approaches, it was still hard to define the degree of correlation and closeness among topics.

One of the statistical methods for handling correlated structure is network analysis. Specifically, network modeling estimates the relationships among nodes based on their dependency structure. Moreover, it can provide global and local representation of nodes at the same time. In this context, \citet{mei2008topic} tried to identify topical communities by combining a topic model with a social network. Specifically, it estimated topic relationships given the known network structure and tried to separate out topics based on connectivity among them. However, in this approach, the network information needs to be provided to construct and visualize the topic network, which is not a trivial task in practice. So, we want to use a network model to simultaneously estimate relationships among topics and visualize their dependency structure.

In addition, we also aim to discover the meaning of topics in two ways. First, we interpret the meaning of the topics by selecting meaningful words that represent the characteristics of the topic. To select the meaningful words within the topics, we provide the score based on the information provided by our analytical framework. According to \citet{airoldi2016improving}, they evaluated the topics using the novel score called \textit{FREX}, which quantifies words' closeness within the topic based on the word frequency and exclusivity. Here, $\textit{FREX}_{i,j}$ reflects how word $j$ is exclusively close to topic $i$ compared to other topics. However, given the short text data, the frequency of words in the documents would not be sufficient to distinguish meaningful words within topics. 
So, we've come up with a score that takes into account both how likely words are to belong to each topic and how close they are to each topic. The above information is inferred by the topic model and network model, respectively. By putting these two pieces of information together, our analytical framework can pick up information about relativeness and exclusivity. Second, we track the change in relationships among topics using different word sets to determine whether the topics' relationships with other topics are changing dramatically or steadily remaining in similar locations.

The rest of this article is organized as follows. In Section 2, we will introduce our methods and describe how we combine two different statistical approaches: (1) topic modeling, specifically the BTM, and (2) the gaussian version of the latent space item response model \citep{Jeon:2021}, which estimates associations between items based on the latent distance between item and response. In Section 3, we will apply our approach to the COVID-19 literature to evaluate and demonstrate the usefulness of our approach. In Section 4, we will summarize our analytical framework and contributions.

\section{Method}

We developed a novel framework for estimating topic relationships through an interaction map, which positions topics on an Euclidean latent space. This framework consists of the following four steps, as illustrated in Figure \ref{analysis_frame}. In the first step, we implement text mining with natural language processing and construct the word corpus by expanding a set of words using the {\tt word2vec} model \citep{Griffiths2004CGS}. In the second step, based on this word corpus, we extract the topic-word distribution, where each topic is characterized with the corresponding topic-specific distribution of words estimated using the BTM. Using this topic-word distribution, we can extract meaningful words that affect topic characteristics. In the third step, the topic relationships are estimated using the latent space item response model \citep{Jeon:2021}, which provides latent topic positions. Finally, in the fourth step, we can map the relationships among topics with topic-specific traces.

\begin{figure}[htbp]
\setlength{\fboxsep}{0pt}%
\setlength{\fboxrule}{0pt}%
\begin{center}
\includegraphics[width = 1.0\textwidth]{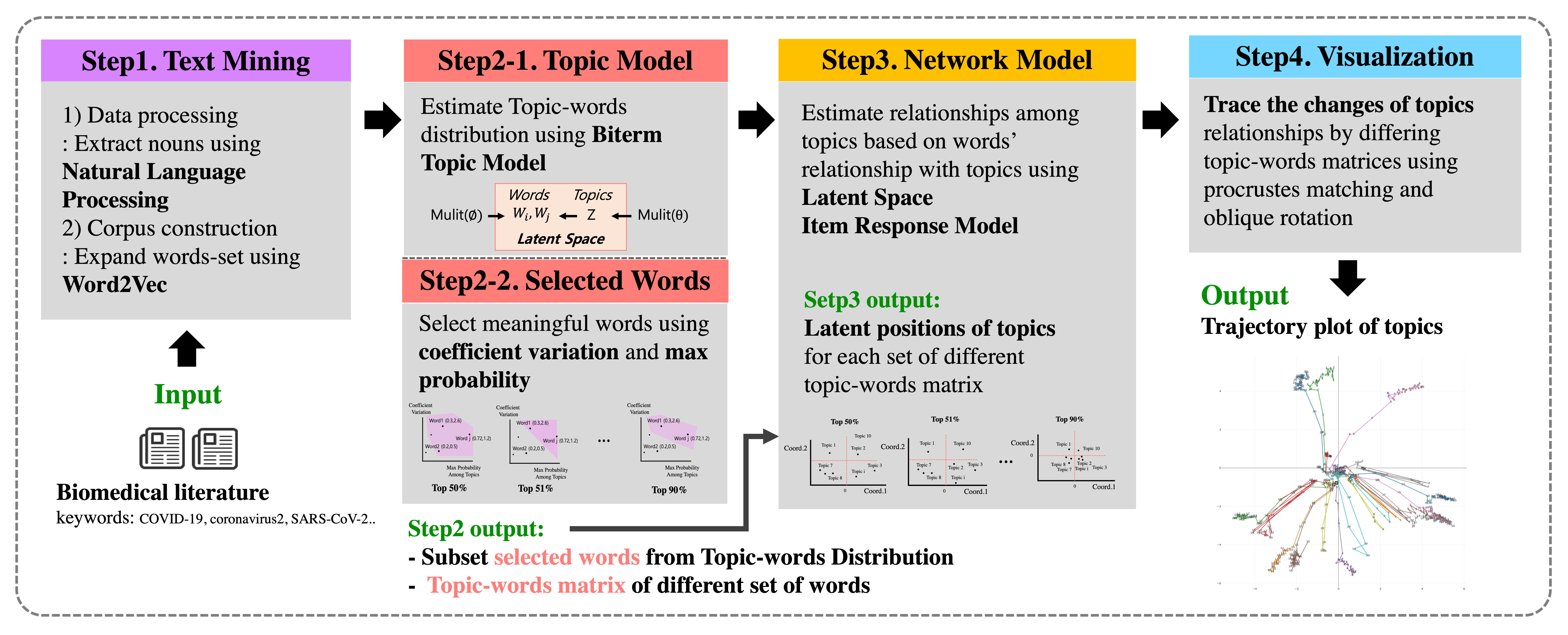}
\end{center}
\caption[]{\textbf{Illustration of the proposed approach for the trajectory topic interaction map} Step 1: Processing of text mining results with natural language processing and construction of the word corpus. Step 2: Extraction of topics, each of which is defined by its own word distribution estimated using the Biterm Topic Model (BTM). Step 3: Estimate the interactions among topics based on their word distributions as networks, using the latent space item response model \citep{Jeon:2021}. Step 4: Visualization of relationships among topics using the trajectory plot showing the transition of topics' coordinates.}
\label{analysis_frame}
\end{figure}

\subsection{Estimating Latent Topics using Biterm Topic Model}
The abstract is an excellent source for understanding the overall text. Since most abstracts are limited to 200 words, they could be regarded as short texts. Therefore, we use BTM for literature mining. To implement BTM, we first need a word corpus, and, then, we extract biterms from each paper, which is input for the BTM. We extract information from the text using morphology analysis, one type of natural language processing technique. Specifically, it splits each word with a suffix to identify the base element unit of a term. Among the basic units of words, we first extract nouns in their most basic forms. This set of nouns is called a corpus. We further expand the corpus by adding relevant words using {\tt word2vec}, which vectorizes distances among words in the Euclidean latent space according to their similarity in meaning. We extract neighboring words based on these distances and enlarge the word sets by collecting these words. Following this, it is necessary to understand the overall semantic structure of a given text. It summarizes the latent structures, called topics, by identifying topics and estimating their corresponding clusters of words. That is, we can estimate topics and their distributions of words using topic modeling.


The likelihood of BTM consists of both topic-word distribution and topic distribution. Therefore, we need two sets of parameters to estimate the topic distribution $\boldsymbol{\theta}$ and the topic-word distribution $\boldsymbol{\phi}_z$. The whole likelihood of BTM is constructed as follows. First, the prior distribution for words ($\boldsymbol{\phi}_{z}$) is set to Dirichlet distribution with hyper-parameter $\beta$ while the prior distribution for topics is set to Dirichlet distribution with hyper-parameter  $\alpha$. Next, we represent the topic with the latent variable $z$, which follows Multinomial distribution with parameter $\theta$. Likewise, each word follows Multinomial distribution with parameter $\boldsymbol{\phi}_z$ so that each word can be generated from a specific topic. Therefore, there are three parameters to estimate, including $\boldsymbol{\phi}_{z}$, $\boldsymbol{\theta}$, and ${z}$. We provide the details of the BTM in Web Appendix A of the Supporting Information. Using the collapsed Gibbs sampling\citep{liu1994collapsed}, we can construct distribution of $\boldsymbol{\phi}_{w|z}$ and $\mathbf{\theta}_z$ with estimated statistics $n_{w|z}$ and $n_z$, given as follows: 
\begin{equation}\label{eq:3}
    \boldsymbol\phi_{w|z} = \frac{n_{w|z}}{\sum_{w}{n_{w|z}}+M\beta}, \quad \mbox{and} \quad 
    \boldsymbol\theta_{z}=\frac{n+\alpha}{|B|+K\alpha}
\end{equation} 

 After that, we can obtain the topic-word distribution $\boldsymbol\phi_{w|z}$ and the topic distribution $\boldsymbol\theta_z$. Each topic contains words and their corresponding probabilities, so it is meaningful to compare each topic based on their word distributions. The simplest way to distinguish topics is to compare word memberships between topics. Since the output of each topic-word distribution includes all words, we might not be able to determine characteristics of topics if we use all the words that topics share. Therefore, we select only meaningful words from each topic to estimate the relationships among topics based on their representative words. 
 
In BTM, we can construct the $\mathbf{X}$ matrix with dimension $N\times P$, where $N$ denotes the number of words, $P$ refers to the number of topics, and each cell represents the probabilities that each word belongs to each topic. Since matrix $\mathbf{X}$ contains a redundant set of words, it is more efficient to select meaningful word sets to estimate the topic interactions. To determine a criterion to choose words, we calculate the coefficient of variation and maximum probability from each row of the matrix $\mathbf{X}$. Each value represents the variation of word probabilities among different topics and the degree of a word's linkage to a particular topic. Since BTM results in the topic-word distribution, where each word's probability is measured within each topic, we need to adjust their variation rather than simply calculating a variance. Therefore, we divide the standard deviation by the mean of the words' probabilities for each topic. 

There are two reasons for choosing a coefficient variation and maximum probability from each row of matrix $\mathbf{X}$ as a criterion to select words. First, it is expected that important words have large variations in probabilities among topics because if a word has low variation across topics, it is likely that the word does not specifically represent any topic. Therefore, meaningful words can be selected based on their variations, and using the coefficient of variation can further scale the words' dispersion among topics. Second, in order to be a meaningful word, it should also have a high probability in at least one topic. For example, if a word has high variation but only low probabilities across topics, it still cannot differentiate topics. Therefore, we can effectively characterize the topic by selecting a word with high probability in at least one topic and those with large variation. 

\subsection{Estimating Topic Relationships and Visualizing the Trace of Interaction Map Using Latent Space Item Response Model} We estimate interactions among topics and visualize their relationships by mapping them on the interaction map. \citet{hoff2002latent} proposed the latent space model, which expresses a relationship between actors of a network in an unobserved ``social space'', so-called latent space. Inspired by \citet{hoff2002latent}, \citet{Jeon:2021} proposed the latent space item response model (LSIRM) that viewed item response as a bipartite network and estimated the interaction between respondents and items using the latent space.

Given our goal to estimate the interactions among topics and visualize their latent positions on the interaction map based on their associated words, we use LSIRM with $\mathbf{X}$ as a bipartite network, where an item indicates a topic and respondents refer to words. However, the original LSIRM proposed by \citet{Jeon:2021} cannot be directly applicable here because it was designed for binary item response dataset, where each cell in the item response data has a binary value (0 or 1). On the contrary, here our input data $\mathbf{X}$ has continuous probabilities indicating how likely each word belongs to each topic. Therefore, in order to apply LSIRM to our input data $\mathbf{X}$, we need to expand the \citet{Jeon:2021} model to a Gaussian version, which is described in detail below. The modified Gaussian version of LSIRM can be written as  
\begin{align*}
    \begin{split}
        x_{i,j} \mid \boldsymbol{\Theta} &= \boldsymbol\beta_i +\boldsymbol\theta_j - || {\bf v}_i - {\bf u}_j || + \epsilon_{i,j}, \\ 
        \epsilon_{j,i} &\sim \mbox{N} (0, \sigma^2),
    \end{split}
    \label{eq:LSIRM}
\end{align*}
where $x_{j,i}$ indicates the probability that word $j$ belongs to topic $i$, for $i = 1, \cdots, P$ and $j = 1, \cdots, N$. Because the original LSIRM use logit link function to handle the binary data, here we use the linearity assumption between $x_{j,i}$ and the attribute part with the interaction part. We add an error term $\epsilon_{j,i} \sim N(0, \sigma^2 )$ to satisfy the normality equation. We use the notation $\boldsymbol{\Theta} = \{\boldsymbol{\beta}=\{\boldsymbol\beta_i\}, \boldsymbol{\theta}=\{\boldsymbol\theta_j\}, \bf{U}= \{{\bf u}_j\},  \bf{V}=\{{\bf v}_i\} \}$, and $|| {\bf v}_i-{\bf u}_j  ||$ represents the Euclidean distance between latent positions of word $j$ and topic $i$. Here, the shorter distance between $\mathbf{v}_i$ and $\mathbf{u}_j$ implies the higher probability that word $j$ links to topic $i$. Therefore, the latent positions of topics can be estimated based on the distances between words. Given the model described above, we use Bayesian inference to estimate parameters in the Gaussian version of LSIRM. We specify prior distributions for the parameters as follows:
\begin{align*}
    \boldsymbol\beta_i \lvert \tau^2_\beta &\sim \text{N}(0,\tau^2_\beta),\quad \tau^2_\beta >0 \\
    \boldsymbol\theta_j \lvert \sigma^2 &\sim \text{N}(0,\sigma_{\theta}^2),\quad \sigma^2 >0 \\\sigma^2 &\sim \text{Inv-Gamma}(a , b),\quad a_>0 , \quad b>0 \\
    \sigma_{\theta}^2 &\sim \text{Inv-Gamma}(a_\sigma , b_\sigma),\quad a_\sigma>0 , \quad b_\sigma>0 \\
    \bf{u}_j &\sim \text{MVN}_d (\mathbf{0, I}_d) \\
    \bf{v}_i &\sim \text{MVN}_d (\mathbf{0, I}_d),
\end{align*}
where $\mathbf{0}$ is a length-$d$ vector of zeros and $\mathbf{I}_d$ is the $d \times d$ identify matrix. We fixed $\tau^2_\beta$ as a constant value. The posterior distribution of LSIRM is 
\begin{equation*}
    \pi(\boldsymbol{\Theta}, \sigma^2 |\bf{X} ) \propto \prod_{j}\prod_{i} \mathbb{P} \left (x_{ji}| \boldsymbol{\Theta} \right) 
    \prod_j \pi(\boldsymbol\theta_j | \sigma_{\theta}^2 ) \pi(\sigma_{\theta}^2) \prod_i \pi(\boldsymbol\beta_i )  
    \prod_j \pi(\bf{u} _j )  \prod_i \pi( \bf{v}_i)  \pi(\sigma^2)
\end{equation*}
and we use Markov Chain Montel Carlo (MCMC) to estimate the parameters of LSIRM. In this way, we can obtain latent positions of $\mathbf{u}_j$ and $\mathbf{v}_i$ on the interaction map $\mathbf{R}^{d}$. Since we are interested in constructing the topic network, we utilize $\mathbf{v}_i, i=1,\cdots, P$ and make it as matrices of ${\bf A} \in \mathbf{R}^{P\times d}$ where row indicates P number of topics and column indicates d number of dimension of coordinates.

 In order to further improve the interpretation of relationships among topics, we trace how their latent positions change as a function of word sets. Specifically, we compare topics' latent positions ${\bf A}_k$ from the various sets of matrices $\mathbf{X}_k, k=1,\cdots,K$ to estimate and trace their interactions based on different word sets. After proceeding with the LSIRM model with various sets of matrices $\mathbf{X}_k$, we can obtain matrices of ${\bf A}_k$, composed of coordinates of each topic. For this purpose, we implement a Procrustes matching two times. First, we implement so-called within-matrix matching within the MCMC samples for each topic's latent positions generated from LSIRM. Note that there exist multiple possible realizations for latent positions because the distances between pairs of respondents and items, which are only included in a likelihood function, are invariant under rotation, translation, and reflection \citep{hoff2002latent, shortreed2006positional}. In order to tackle this invariance property for determining latent positions, we implement within-matrix Procrustes matching \citep{borg2005modern} as post-processing of MCMC samples. 

Second, we implement so-called between-matrix matching for the estimated matrices to locate topics in the same quadrant. To align the latent positions of topics, we need to set up the baseline matrix, denoted as ${\bf A}_\text{max}$, which maximizes the dependency structure among topics. To measure the degree of dependency structure,  we take the average of the distances of topics' latent positions from the origin. The longer distance of latent positions from the origin implies a stronger dependency on the network. It helps nicely show the change of topics' latent positions because those rotated positions ${\bf A}_k$ from each matrix ${\bf X}_k$ are based on the most stretched-out network from the origin. As a result, we can obtain the repositioned matrices $\mathbf{A}^*_k$, which still maintain the dependency structure among topics but are located in the same quadrant.

With the oblique rotation, the interpretability of axes can be further improved and topics can be categorized based on these axes. For this purpose, we apply the {\tt oblim} rotation \citep{jennrich2002simple} to the estimated topic position matrix $\mathbf{A}^*_k$, using the R package {\tt GPAroation} \citep{bernaards2005gradient}. We denote the rotated topic position metric by $\mathbf{B}_k$. To interpret the trajectory plot showing traces of topics' latent positions, we extract the rotation information matrix ($\mathbf{R}$) resulting from an oblique rotation as the baseline matrix $\mathbf{B}_\text{base}$. Then, we multiply each matrix ($\mathbf{B}_k$) by the rotation matrix ($\mathbf{R}$) to plot the topics' latent positions. 

\subsection{Scoring the Words Relation to Topics}

By combining the idea from \citet{airoldi2016improving} with the interaction information among topics in our problem, here we propose the score $s_{i,j}$ which measures the exclusiveness of word $j$ in the topic $i$. We define score $s_{i,j}$ as 
\begin{equation}
    \begin{split}\label{eq:score}
        s_{i,j} &= \left( \frac{w_1}{2-\text{ECDF}_{\boldsymbol\delta_{.,j}}(\delta_{i,j})} \right)
        +\left(\frac{w_2}{2-\text{ECDF}_{\boldsymbol\delta_{i,.}}(\delta_{i,j})} \right)\\
        &+\left(\frac{w_3}{1+\text{ECDF}_{\boldsymbol\gamma_{.,j}}(\gamma_{i,j})} \right)
        +\left(\frac{w_4}{1+\text{ECDF}_{\boldsymbol\gamma_{i,.}}(\gamma_{i,j})} \right) ,
    \end{split}
\end{equation}
where $w_1,w_2,w_3,$ and $w_4$ are the weights for exclusivity (here, we set $w_1=w_2=w_3=w_4=0.25$) and $\text{ECDF}$ is the empirical CDF function. Here, $\delta_{i,j}$ denotes the probability of word $j$ belonging to topic $i$ given by BTM. On the other hand, $\gamma_{i,j}$ is the distance between the latent position of word $j$ and topic $i$ estimated by LSIRM. The higher value of $\delta_{i,j}$ corresponds to the closer relationship between word $j$ and topic $i$. On the other hand, the smaller value of $\gamma_{i,j}$ indicates the shorter distance between word $j$ and topic $i$. To make the meaning of the shorter distance and the higher probability consistent as both contribute higher scores, we subtract $\text{ECDF}_{\boldsymbol\delta_{.,j}}(\delta_{i,j})$ and $\text{ECDF}_{\boldsymbol\delta_{i,.}}(\delta_{i,j})$ from 2. Note that here we use 2 to avoid the zero in the denominator. For example, if $\delta_{i,j}$ has a high probability within the topic $i$ and between the other topics and word $j$, then both $2-\text{ECDF}_{\boldsymbol\delta_{.,j}(\delta_{i,j})}$ and $2-\text{ECDF}_{\boldsymbol\delta_{i,.}(\delta_{i,j})}$ will have small values. This means that the word $j$ is distinctive enough to represent the meaning of topic $i$. Likewise, if the latent distance between word $j$ and topic $i$ has the shortest distance within topic $i$ than between the other topics and word $j$, then the $1+\text{ECDF}_{\boldsymbol\gamma_{.,j}(\gamma_{i,j})}$ and $1+\text{ECDF}_{\boldsymbol\gamma_{i,.}(\gamma_{i,j})}$ have the smallest value contributing to a high score $s_{i,j}$. Here, we add 1 to each denominator term, preventing the denominator from becoming zero. Based on $s_{i,j}$, we can determine whether the word $j$ and topic $i$ are close enough to be mentioned in the same document. By collecting the high-score words, we can characterize the topics and name them.

\section{Results}
We applied the proposed approach to the rapidly increasing COVID-19 literature to investigate its latent semantic structure at the beginning of the pandemic. Specifically, we downloaded the COVID-19 articles published from the PubMed database (\url{https://pubmed.ncbi.nlm.nih.gov}) with a timeframe between 1st, 2019, and August 3rd, 2020, coinciding with the date of the WHO's designation of COVID-19 as a pandemic.} Specifically, we collected articles whose titles contain ``coronavirus2'', ``covid-19'' or ``SARS-CoV-2'' and this approach resulted in a total of 35,585 articles. After eliminating articles without abstracts (i.e., only titles or abstract keywords), our final text data contained a total of 15,015 documents. 

To construct the corpus, we used abstract keywords that concisely captured the messages delivered by the paper. To achieve the richer corpus, we also used the {\tt word2vec} \citep{mikolov2013efficient} to train against relationships between nouns from the abstract and the abstract keywords. Specifically, {\tt word2vec} extracted nouns from abstracts, which were embedded near the abstract keywords, and added those selected words to the corpus. 

Using the trained {\tt word2vec} network with 256 dimensions, we selected ten words from the abstract nouns that were near each abstract keyword. We provide the details of the training {\tt word2vec} network in the Web Appendix B of the Supporting Information. The corpus construction results in 9,643 words from 15,015 documents. We further filtered out noise words, including single alphabets, numbers, and other words that are not meaningful, e.g., `p.001', `p.05', `n1427', `l.', and `ie'. Finally, to obtain more meaningful topics, we removed common words like `data', `analysis', `fact', and `disease'.

To implement BTM, we set the topic number to 20. For the hyper-parameters, we assigned $\alpha$ = 3 and $\beta$ = 0.01. Since our main goal is to visualize the topic relationships, we empirically searched and determined the hyper-parameters. The posterior distribution of the topic-word was estimated using the Gibbs sampler. Specifically, we generated samples with 50,000 iterations after the 20,000 burn-in iterations and then implemented thinning for every $100^{\text{th}}$ iteration. In each topic-word distribution obtained from BTM, words with high probabilities characterize the topic. 

Histograms of log-transformed probabilities show bimodal topic-word distributions. This pattern indicates that there are some words that had low probabilities of belonging to a specific topic, whereas the mode in the center corresponds to the words that have high probabilities enough to characterize the meaning of the topic. We provide the histograms of various topics in Web Appendix D of the Supporting Information. Therefore, it might be more desirable to estimate topic relationships using only the words corresponding to the mode in the center rather than all the words. We checked the minimum cutoff value based on these histograms, which ranged between -11 and -12. We calculated the number of words whose log-scaled probabilities were above -11 to -12 to specify the minimum number of words for estimating the topic network. We found that more than 1,000 words are needed to represent topics properly. Based on this rationale, we decided to use at least 1,000 words to estimate the positions of topics in the latent space based on the positions of words. Given the selected minimum number of words, we extract meaningful words that can discriminate characteristics of topics based on the coefficient of variation and maximum probability.

In this study, rather than using a fixed number of words, we investigated relationships among topics identified with different numbers of words, which were determined using the two criteria described above. Specifically, we obtained multiple matrices corresponding to the top 60\% to 40\% of words determined based on the two criteria. The numbers of words corresponding to the 60-th and 40-th percentiles were 2,648 and 1,095, respectively. Therefore, we used 21 sets of matrices, $\mathbf{X}_k (k = 40\%, \cdots, 60\%)$, as the LSIRM input data, where their dimensions were ranged from $2,648 \times 20$ to $1,095 \times 20$. 

In this way, we obtained the 21 sets of matrices ${\bf X}_k$ and we considered 20 topics for all the matrix sets. To estimate topics' latent positions $\bf{V}= \{{\bf v }_i\}$ where $i=1,\cdots,20$, MCMC was implemented. The MCMC ran 55,000 iterations, and the first 5,000 iterations were discarded as burned-in processes. Then, from the remaining 50,000 iterations, we collected 10,000 samples using a thinning of 5. To visualize relationships among topics, we used two-dimensional Euclidean space. Additionally, we set 0.28 for $\boldsymbol\beta$ jumping rule, 1 for $\boldsymbol\theta$ jumping rule, and 0.06 for ${\bf w}_j$ and ${\bf z}_i$ jumping rules. Here, we fixed prior $\boldsymbol\beta$ and $\boldsymbol\theta$ follow N(0,1). We set $a_{\sigma}=b_{\sigma}=0.001$. 

LSIRM takes each matrix ${\bf X}_k$ as input and provides the ${\bf A}_k$ matrix as output after the Procrustes-matching within the model. Since we calculate topics' distance on the 2-dimensional Euclidean space, ${\bf A}_k$ is of dimension $20 \times 2$. We visualized interactions among topics using the baseline matrix ${\bf A}_{\max}$ chosen so that we can compare topics' latent positions without having identifiability issues from the invariance property. From ${\bf A}_k$, we calculated the distance between the origin and each topic's coordinates. The closer distance of a topic position from the origin indicates the weaker dependency with other topics. 

The dependency structure among topics starts to be built up from ${\bf A}_{47\%}$. There are two possibilities that can lead to low dependency. First, it is possible that a small number of words could distinguish the characteristics of their topics from the other topics. Second, it is possible that most of the words were commonly shared with other topics. We provide the distance plot of ${\bf A}_k$ in the Web Appendix E of the Supporting Information.
Based on this rationale, we chose ${\bf A}_{47\%}$ as the baseline matrix ${\bf A}_{\max}$. With this baseline matrix ${\bf A}_{47\%}$, we implemented the Procrustes matching to align the direction of the topic's latent positions from each matrix ${\bf A}_k$. Using this process, we could obtain the ${\bf A}_k^{*}$ matrix matched to the baseline matrix ${\bf A}_{47\%}$. We named the identified topics based on top ranking words using the ${\bf A}_{47\%}$ matrix. This is because the baseline matrix ${\bf A}_{47\%}$ has the most substantial dependency structure comparing other $\mathbf{A}^*_k$ matrix containing the words that characterize topics nicely. 

We applied {\tt oblim} rotation to the estimated topic position matrix $\mathbf{A}_{k\%}^{*}$ using the R package GPArotation \citep{bernaards2005gradient}, and obtained matrix $\mathbf{B}_k$ $k=40\%, \cdots, 60\%$ with the rotation matrix $\mathbf{R}$.
By rotating the original latent space, the axes improve interpretation of the identified latent space (e.g., determining the meaning of a topic's transition based on the X-axis or Y-axis).

\begin{figure}[htbp]
\begin{subfigure}{.5\textwidth}
  \centering
  \includegraphics[width=0.9\linewidth]{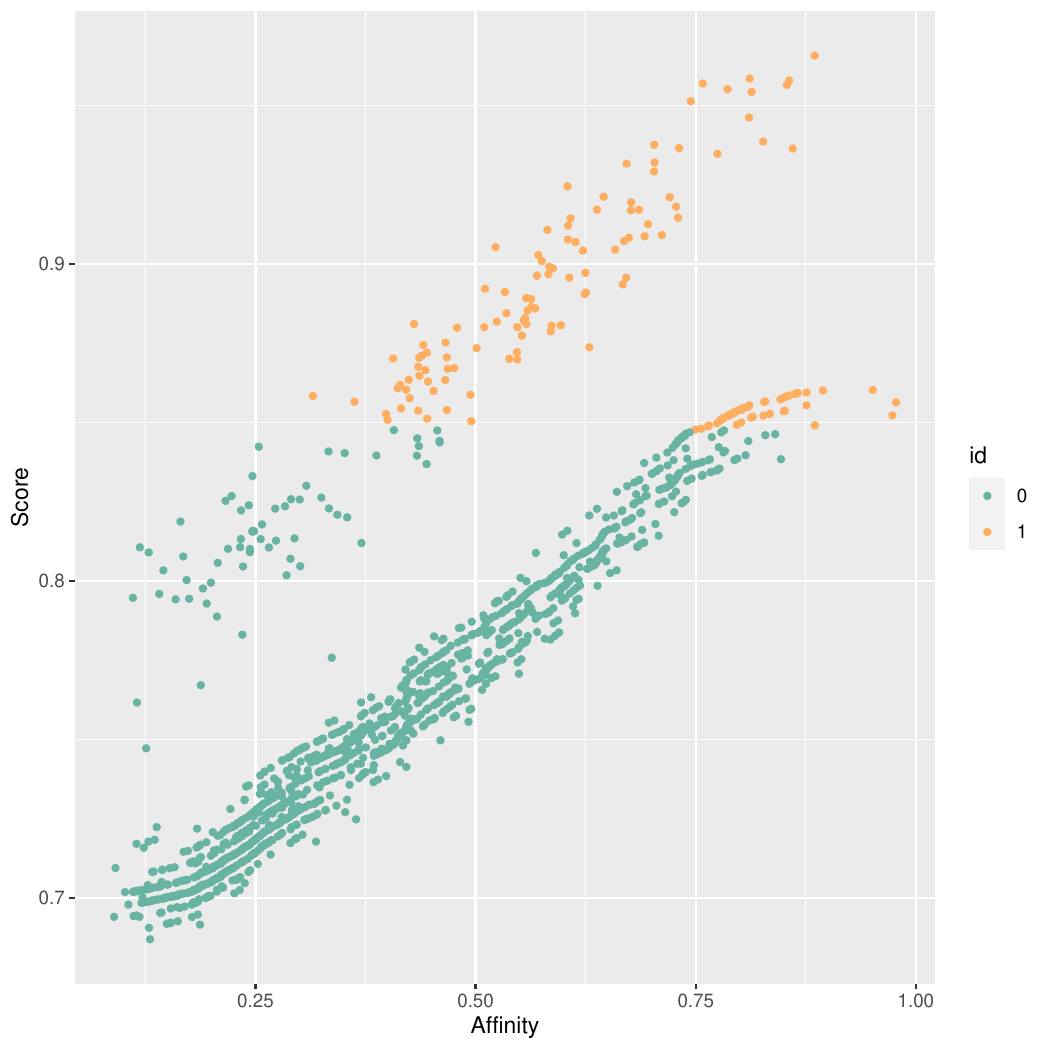}
  \caption{$s_{19,j}$ and affinity plot of Topic 19 in $\mathbf{B}_{47\%}$}
  \label{scorenodepen}
\end{subfigure}%
\begin{subfigure}{.5\textwidth}
  \centering
  \includegraphics[width=0.9\linewidth]{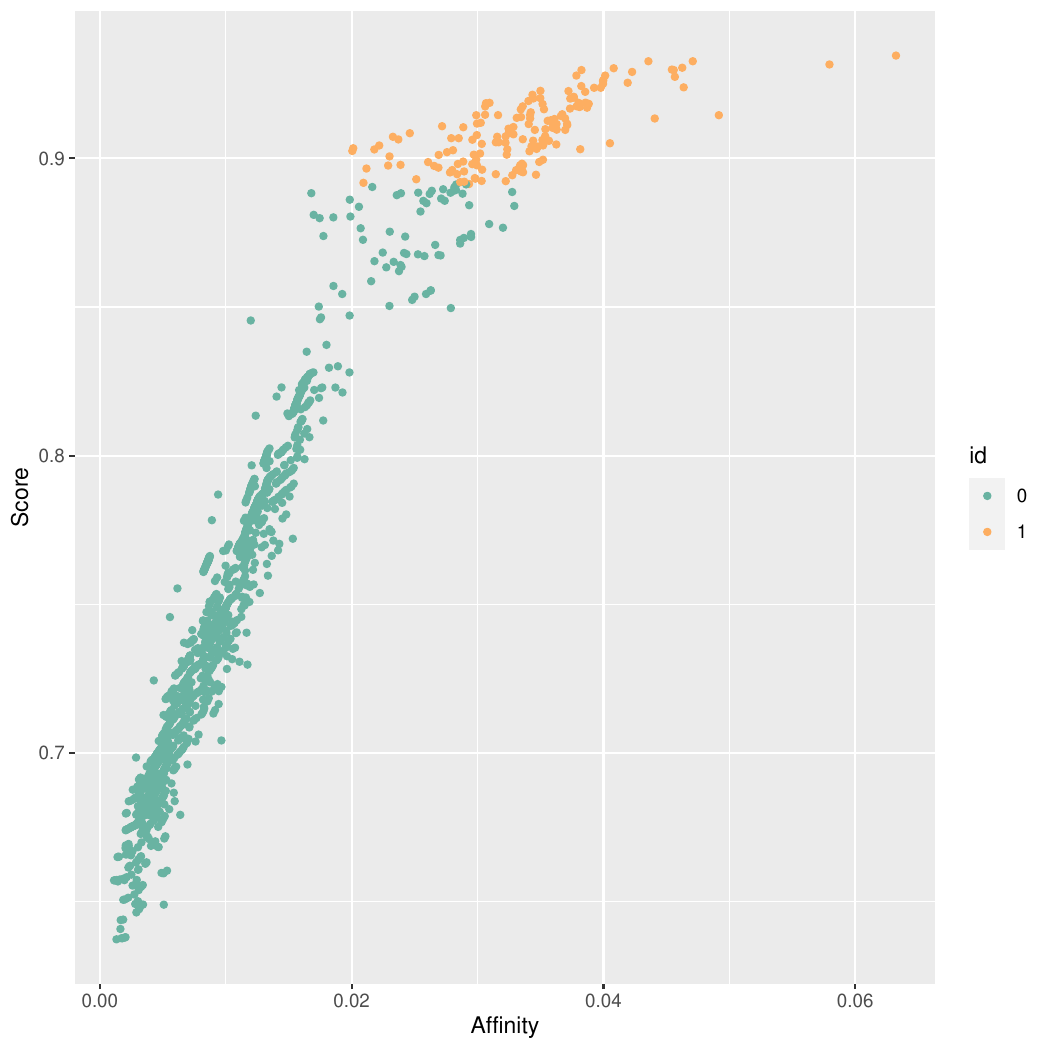}
  \caption{$s_{3,j}$ and affinity plot of Topic 3 in $\mathbf{B}_{47\%}$}
  \label{scoreyesdepen}
\end{subfigure}
\caption{(a) The plot of Topic 19 in ${\bf B}_{47\%}$ has many common words shared by other near-located topics. Some words are close to that topic, but not enough to distinguish the characteristics of that topic from the given information in BTM. (b) The plot of Topic 3 in ${\bf B}_{47\%}$ which is located outside the center. It has more distinctive characteristics compared to other topics.}
\label{scoreplotcompare}
\end{figure}

We calculated the score ${s}_{i,j}$ for each word $j$ with topic $i$ based on each $\bf{B}_{k}$ for $k=40\%, \cdots, 60\%$. Figure \ref{scoreplotcompare} shows the two different scenarios of the plot of $s_{i,j}$ and affinity ($\exp{(-\gamma_{i,j})}$) in ${\bf B}_{47\%}$; one with a high score and high affinity, and the other with a low score but high affinity. Because this topic has many words in common with other nearby topics, although there are some words close to that topic, they were not close enough to distinguish the characteristics of that topic based on the probability of words belonging to it (Figure \ref{scorenodepen}). The second scenario is that the latent positions of some topics are located far from the center, indicating that they have distinctive characteristics compared to other topics. There is only one pattern in this case; it has a high score and affinity because those words are both close to those topics and have a high probability within those topics.

We extracted the top 20\% of words with high values in score to name the topic. After then we collected the meaningful words that commonly appear on the top of the list among every portion of words set ${\bf B}_{k}, k=40\%, \cdots, 60\%$. Table~\ref{topicname} shows the name of each topic determined based on the selected words with its top score words. We provide a detail explanation of the topic name with high-score words in the Web Appendix F of the Supporting Information.

\begin{table}[tt]
\centering
\caption{\label{topicname} \textbf{Topic names based on $\mathbf{A}^{*}_{47\%}$ matrix.} The abbreviation is marked by an asterisk, and the full form can be found in the Web Appendix H of the Supporting Information}
\resizebox{\columnwidth}{!}{%
\begin{tabular}{lllll}
  \hline
Topic  & Name & Top score words  \\
  \hline
1 & Lung Scan & subpleural  & crazy paving   & bronchogram\\ 
 & & (0.963) & (0.957) & (0.950)\\
2 & Compound and Drug & Protein Data Bank &  papain-like & intermolecular \\
& & (0.938) & (0.936) &  (0.934) \\ 
3 & Bacteria, Nano, and Diet & $\text{DHA}^{\ast}$ & biofilm  & vancomycin-resistant  \\
&& (0.968) & (0.968) & (0.967) \\ 
4 & Treatment of Other Diseases & sarcoma &  arthroplasty & neoadjuvant \\
&& (0.948) & (0.944) & (0.943)  \\
5 & Symptoms (Cardiovascular) & vein & antithrombotic & $\text{VTE}^{\ast}$  \\
&& (0.906) & (0.899) & (0.898)  \\ 
6 & Molecular-level Response to Infection & $\text{NLRP3}^{\ast}$ & metalloproteinase & upregulation  \\
&& (0.929) &(0.927) &(0.926) \\ 
7 & COVID-19 Risk Prediction Markers & alanin & $\text{BUN}^{\ast}$ & prealbumin \\
&& (0.976) & (0.972) &(0.969)\\ 
8 & Literature Review & Prospero DB &  Wanfang DB  & meta-analys  \\
 && (0.967) & (0.966) & (0.944)\\
9 & Symptom and Comorbidity & petechiae & guillain-barre syndrome & ageusia\\
&&  (0.958) &  (0.955) & (0.952) \\ 
10 & Cardiovascular & infarction & thromboprophylaxis & $\text{VTE}^{\ast}$ \\
&& (0.912) &(0.909) &  (0.909) \\  
11 & Social Impact & classroom  & pedagogy& zoom  \\
&&  (0.925) & (0.911)  &(0.906) \\ 
12 & Financial and Economical Impact & macroeconomics & monetary & profit \\
&& (0.929) &(0.928) & (0.928) \\ 
13 & Statistical Modeling & Weibull& least-square & $\text{MAE}^{\ast}$\\ 
&& (0.984) & (0.975) & (0.966) \\
14 & COVID-19 Test & $\text{LFIAs}^{\ast}$  & transcription-$\text{PCR}^{\ast}$ & Wantai \\
&&  (0.979) & (0.974) &(0.972) \\ 
15 & Psychological/ Mental Issues & anxious&$\text{PTSD}^{\ast}$& post-traumatic \\
&& (0.942) & (0.940) &  (0.935)\\
16 & Lung Scan &  procalcitonin & ground-glass & opacity \\
&& (0.924) &  (0.922) &(0.916) \\
17 & Treatment & reintubation &  multi-centric& Hydroxychloroquine \\
&& (0.943) & (0.943) &  (0.937) \\
18 & Air, Mask, Breathing & laryngoscope& facepiece & supraglottic \\
&& (0.978) & (0.978) & (0.975)\\ 
19 & Prevention of COVID-19 & alcohol-based & decontamination &rub\\
&& (0.944) &  (0.941) & (0.919)\\ 
20 & Immunology and Virus &  $\text{MHC}^{\ast}$ &  multi-epitope & immunodominant \\
&& (0.976) & (0.975) &(0.973)  \\ 
\hline
\end{tabular} %
}
\end{table}

\begin{figure}[htbp]
\centering
\includegraphics[width = 0.9\linewidth]{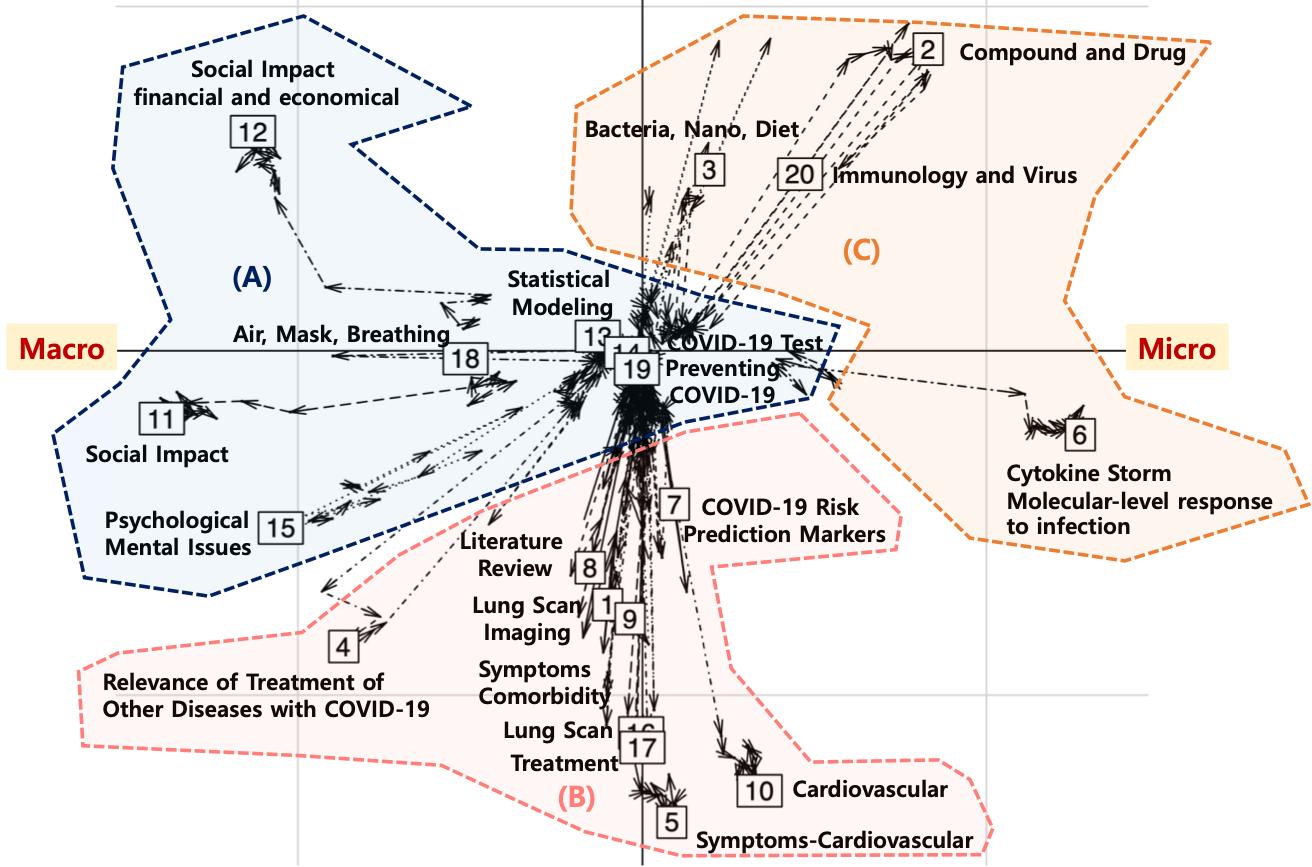}
\caption[]{Trajectory interaction topic map with three topic clusters (A) - (C). (A) About COVID-19 and its effects on different areas. (B) Symptoms and treatment of COVID-19, and its relative studies, works, and  (C) Molecular studies associated with the COVID-19 infection.}
\label{finalresults}
\end{figure}

To eliminate visualization bias due to the selection of words, we tracked the topics' latent positions by using different input matrices $\mathbf{B}_{k}$. We interrogated what kinds of topics have been extensively studied in the biomedical literature on COVID-19. In addition, we also studied how those topics were related to each other, based on their closeness in the sense of latent positions. We also partially clustered the topics based on their relationships using the quadrants. This allows us to check which studies about COVID-19 are relevant to each other and could be integrated. Figure \ref{finalresults} displays the trajectory plot and it shows how topics were positioned on the latent space and how these topics make the transition. More specifically, the direction of arrows refers to how topics' coordinates changed as a function of the numbers of words, where each arrow moved from ${\bf B}_{60\%}$ to ${\bf B}_{40\%}$ as the number of words decreased. 

According to Figure \ref{finalresults}, we observed two distinct groups; the one topic group does not depend on other topics, and the other group shows a dependent relationship with other topics. The former group consists of topics that have common words that are shared by other topics as well. On the other hand, the latter group consists of topics that have their own distinct words. In the former group, the topics `Statistical Modeling', `COVID-19 Test', and `Preventing COVID-19' (Topics 13, 14, and 19, respectively) were located in the center of the plot. This indicates that no matter how many words were used to estimate the topics' latent positions, those topics remained as general topics and have shared many words with other topics. This makes sense given the fact that many articles have mentioned the outbreak of COVID-19, the testing of COIVD-19, and various ways of studying these notorious pandemics through statistical models. For example, among the 137K publications mentioning ``COVID-19'' in the PubMed database, more than 76,000, 64,782 ,and 44,602 publications also mentioned ``outbreak'', ``testing'', and ``prevention'' respectively. The latter group, the topics `Social Impact (financial and economical)', `Social Impact', `Cardiovascular', and `Cytokine Storm' (Topics 12, 11, 10, and 6, respectively) were located away from the center in the plot, which implies their dependency structures with other topics. These topics usually stay on the boundary of the plot regardless of the number of words because they consist mainly of unique words. Finally, topics like `Psychological Mental Issues', `Relevance of Treatment', and `Compound and Drug' (Topics 15, 4, and 2, respectively) start from the origin, move away from the origin for a while, and then return to the origin. This implies that it could not maintain the nature of the topic when fewer words were considered, and it is likely that those topics are either ongoing research or burgeoning topics that have not been studied enough yet.

In addition, we interpreted the topics' meaning based on their latent positions. We now make interpretations using subsets of topics divided by directions. Since we implemented the oblique rotation that maximizes each axis' distinct meaning, we can render meaning to a direction. Figure \ref{finalresults} indicates that there are three topic clusters. First, the center cluster denoted as (A) in Figure \ref{finalresults} are about the outbreak of COVID-19 and its effects on different areas, including the outbreak of COVID-19, diagnosis or testing of COVID-19 using statistical models, the effects on general social, financial, or economic problems, and the mental pain of facing the pandemic shock of COVID-19. Second, the topics located on the bottom side of the plot (cluster (B) in Figure \ref{finalresults}) are related to the symptoms of COVID-19 and their treatment. For instance, there are studies of COVID-19 relating to their symptoms, such as cardiovascular diseases, with lung scan images. On the other side, there are topics related to treatment based on symptoms relevant to COVID-19 and risk prediction markers that validate the treatment effects. These subjects pertain to  `Literature Review (8)', `Lung Scan Imaging (1)', `Symptoms Comorbidity (9),' `Lung Scan (16)', `Treatment (17)', `Relevance of Treatment (4)', `COVID-19 Risk Prediction Markers (7)', and `Cardiovascular (10)'. Finally, the cluster (C) is related to what happens inside our body in response to COVID-19, e.g., cytokine storm, immune system responding to the COVID-19 infection, and the compounding of drugs that investigate the mechanism of SARS-CoV-2. Note that we can also interpret each axis. For example, the x-axis can reflect a spectrum from macro (more negative) to micro perspectives (more positive).

In summary, we identified three main groups of the COVID-19 literature; the outbreak of COVID-19 and its effects on society, the studies of symptoms and treatment of COVID-19, and the COVID-19 effect on the body, including molecular changes caused by COVID-19 infection. We can derive another insight from the locations of clusters. Specifically, from Cluster (A) to Cluster (C), we can observe a counter-clockwise transition from macro perspectives to micro perspectives. Specifically, this flow starts with the center cluster (A) related to the occurrence of COVID-19 and the social impact of COVID-19, followed by studies of the symptoms and treatment of COVID-19 (cluster (B)) and then ends with the clusters (C), which are related to micro-level events, e.g., how SARS-CoV-2 binding occurs and how the immune system responds to upon the COVID-19 infection.

\section{Discussion}

In this manuscript, we propose a novel analytical framework to estimate and visualize relationships among topics based on the text mining results. It allows enhancing our understanding of COVID-19 knowledge reported in the biomedical literature, by evaluating topics' networks through their latent positions estimated based on topic sharing patterns. The proposed approach overcame the limitations of existing approaches, especially discrete and static visualizations of relationships among topics.

First, because we position topics in a latent space, relationships among topics can be intuitively investigated and also easy to interpret. This embedded derivation of topic relationships will reduce the burden on data analysts because it does not require prior knowledge about relationships between topics, e.g., a connectivity matrix. Instead, our approach utilizes the topic-word matrix for calculating the latent positions of each topic, where each cell corresponds to the probability of a word's priority linkage to a topic. Since our approach uses continuous values of Gaussian for the modeling purposes, we developed the continuous Gaussian version of LSIRM. It can quantify closeness between topics by reflecting degrees of linking of words to topics, and, hence, using latent positions for topics, we could represent relationships among topics more precisely.

Second, we provide the score $s_{i,j}$ that can define the name of topics without having expert knowledge. Since estimated probability from topic models does not reflect the exclusivity of words toward topics, it may cause misunderstanding of topics. Moreover, it requires expert knowledge to figure out meaningful words for each topic and characterize the topic using these words. Our approach makes it plausible to get exclusiveness of word $j$ with topic $i$ without subjective interpretation. Discarding redundant information using a score helps to figure out the true meanings of each topic. Moreover, our score $s_{i,j}$ assists in evaluating the performance of our approach through how top-scored words from each topic are relevant in context.

Third, we visualize the topic relationship as a trajectory plot to detect the change of interactions of topics over different set of words. This feature has two important properties: (1) it could measure the main location of the topic, which is steadily positioned in a similar place in spite of differing network structures; and (2) we could distinguish popular topics mentioned across articles from recently emerging topics by scanning the latent position of each topic. Specifically, if a particular topic shares most of the words with other topics, it is more likely to be located in the center on the latent interaction map. In contrast, if a specific topic consists mostly of words unique to that topic (e.g., a rare topic or an independent topic containing its own referring words), it is more likely to be located away from the center on the interaction map. For example, in the context of COVID-19, it is more likely that common subjects like `outbreak' and `diagnosis' are located in the center, while more specific subjects like `Cytokine Storm' are located more outside on the interaction map. 

Finally, this approach helps organize a tremendous amount of literature and mine underlying relationships among topics based on the literature. The topic network visualizes the relationship's topics in an intuitive way, which can assist researchers in designing their studies. For instance, if some researchers want to investigate a specific topic, say `COVID-19 symptoms', our framework can assist them by providing information answering the three following questions: (1) Which set of words are associated with `COVID-19 symptoms'? Researchers can obtain this information from the BTM results. In addition, since we extract meaningful words that distinctively represent each topic's meaning, our approach could further support this investigation with more refined word sets. (2) Are there any other topics related to `COVID-19 symptoms', which can be used for extending and elaborating research? Researchers can answer this question by intuitively interrogating the final visualization. (3) Is `COVID-19 symptoms' a common or specific topic? Since we could trace change of latent topic positions as a function of words set, our method provides relevant insights through topic locations on the interaction map. Specifically, researchers can consider `COVID-19 symptoms' to be common if it is located in the center, and to be specific if it is located away from the center. 

The proposed framework can still be further improved in several ways, especially by allowing word-level inference, i.e., extraction of meaningful words that characterize each topic. Although the distance between each topic and relevant words is taken into account in our model when estimating topics' latent positions, simultaneous representation and visualization of words are still not embedded in the current framework. We believe that adopting a variable selection procedure to determine key words can potentially address this issue and this will be an interesting future research avenue.

\section*{Acknowledgements}

This work was supported by the National Institutes of Health [grant numbers R21-HG012482, R21-CA209848, U01-DA045300, US4-AG075931 awarded to DC], Yonsei University Research Fund [grant number 2019-22-0210 awarded to IHJ] and the National Research Foundation of Korea [grant number NRF 2020R1A2C1A01009881; Basic Science Research Program awarded to IHJ]. The funders had no role in study design, data collection and analysis, decision to publish, or preparation of the manuscript.

\bibliography{reference}

\label{lastpage}

\section*{Supporting Information}
Web Appendices are available with this paper at the Biometrics website on Wiley Online Library. The source code and data can be downloaded from https://github.com/jeon9677/gViz.

\vspace*{-8pt}

\end{document}